\documentclass[onecolumn,aps,12pt]{revtex4}
\usepackage{amsmath,graphicx}
\usepackage{amsmath,graphicx} 
\def\,{\ifmmode\mskip\thinmuskip\else\leavevmode\thinspace\fi}

\newcommand{\dd}{\mbox{d}}
\newcommand\ba{\begin{eqnarray}}
\newcommand\ea{\end{eqnarray}}

\def\fun#1#2{\lower3.6pt\vbox{\baselineskip0pt\lineskip.9pt
\ialign{$\mathsurround=0pt#1\hfil##\hfil$\crcr#2\crcr\sim\crcr}}}

\begin{document}

\title{Radiative corrections to DVCS electron tensor}

\author{V.V.~Bytev}
\affiliation{Joint Institute for Nuclear Research, 141980 Dubna,
Russia}
\author{E.A.~Kuraev}
\affiliation{Joint Institute for Nuclear Research, 141980 Dubna,
Russia}

\date{\today}

\begin{abstract}
We had considered RC to the virtual Compton scattering
in the high-energy limit including additional hard photon
emission in leading logarithmical approximation. Our 
result is consistent with Drell-Yan picture of the process
and expressed in terms of electron structure function.
The comparison with previous work on VCS is done.
\end{abstract}

\maketitle

\section{Introduction}
\label{sect1}

The QED radiative corrections (RC) to virtual Compton scattering 
($ep\to ep\gamma$) was calculated in lowest order in 
\cite{MVand}, where the detailed study of one-loop virtual corrections
including first-order soft photon emission contribution was done.
Besides that, in that work the radiative tail corresponding to
the case of hard photon for collinear kinematics was considered, 
and results were compared with experimental data.

Results, obtained in our paper (virtual and soft photon emission corrections) 
in high energy limit are in agreement with the results of \cite{MVand}.
Nevertheless, we believe that investigation of RC was not done 
completely in \cite{MVand}: the emission of additional hard photon
was not considered as well as estimation of higher order contributions.
This is motivation of our paper.

The paper organized as follows:
In part 2 we consider Born approximation of the process and relevant kinematics.
In 3 part we consider the contributions of three gauge invariant
classes of 1-loop virtual corrections. In part 4, 5 the soft and additional
hard photon emission in collinear kinematics is considered.
Aso we take into account the Dirac form-factor contribution.
In part 6 we reveal the form of RC in leading logarithmical 
approximation, which is consistent with renormalization group approach
and perform the relevant generalization for all orders in 
leading logarithmical approximation (LLA)
in forms of electron structure functions.
In conclusion we consider some realistic model for quark-quark 
scattering with hard photon emission.
The value of $K$-factor can be inferred in principle, combining the results 
of \cite{MVand} and nonleading contributions arising from $2$ hard
photon emission corrections.

We gave the explicit expression for DVCS contribution for the case
$e\mu\to e\mu\gamma$ scattering, which can be considered as a model for 
$ep\to ep\gamma$ process.

\section{Born approximation}

We specify the kinematics of the virtual Compton scattering process
\begin{gather} 
e_-(p_-)+p_-(q_-)\to e_-(p_-^{'})+p_-(q_-^{'})+\gamma(k_1),
\end{gather}
as follows:
\begin{gather}
\qquad p_-^2=m^2,\quad p_-^{'2}=m^2, \quad k_1^2=0,
\nonumber \\
\chi_-=2k_1\cdot p_-, \quad \chi_-'=2k_1\cdot p_-^{'},\quad
\tilde{\chi}_-=2k_1\cdot q_-, \quad \tilde{\chi}_-'=2k_1\cdot q_-^{'}, \nonumber \\
s=(p_-+q_-)^2,\quad s_1=(p_-^{'}+q_-^{'})^2,\quad
t=-q^2=-2p_-\cdot p_-^{'}, \\ \nonumber
 t_1=-q_1^2=-2q_-\cdot q_-^{'},\quad u=-2p_-\cdot p_-^{'},\quad
u_1=-2p_-^{'}\cdot q_-,
\end{gather}
where $m$ is the electron mass.
Throughout the paper we will suppose
\begin{eqnarray}
s\sim-t\sim-t_1\sim-u\sim-u_1\sim\chi_-\sim\chi_-^{'}\gg m^2.
\end{eqnarray}

\begin{figure} 
\includegraphics[scale=1]{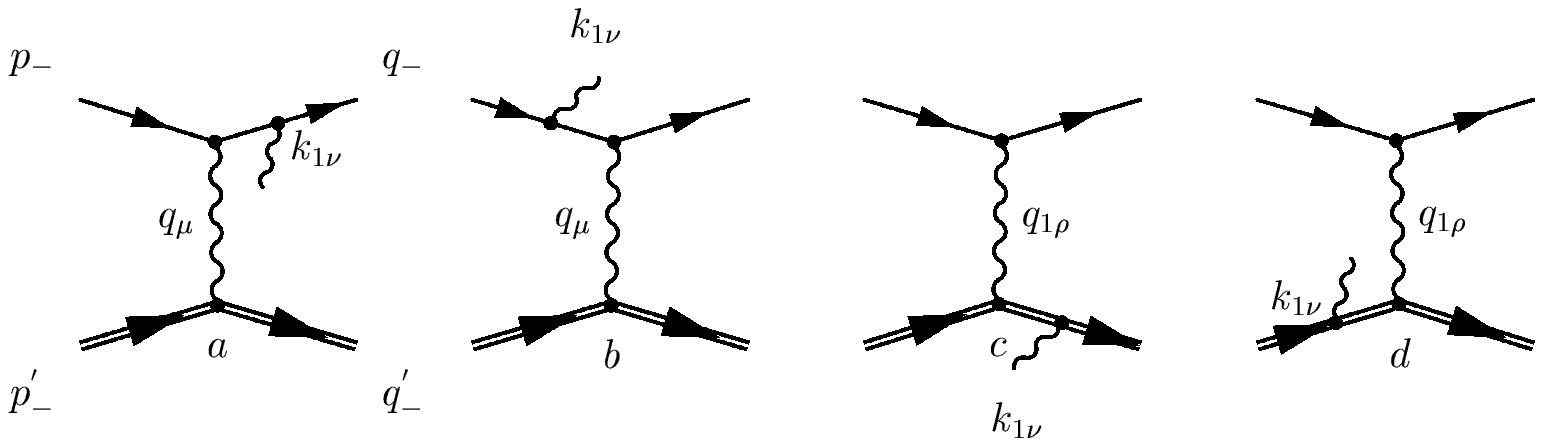}
\caption{Born FD for virtual Compton scattering.}
\label{fig1}
\end{figure}

The value of electronic DVCS tensor in Born approximation reads (see fig.\ref{fig1}a,b):
\begin{gather}
E^{born}_{\mu\nu\rho}(p_-,k_1,p_-^{'})=
Sp\,\, \hat{p}_-^{'}(Q_\nu\hat{\gamma}_\mu+\frac{\hat{\gamma}_\mu\hat{k_1}\hat{\gamma}_\nu}{\chi_-}
+\frac{\hat{\gamma}_\nu\hat{k_1}\hat{\gamma}_\mu}{\chi_-^{'}})\hat{p}_-\hat{\gamma}_\rho\, ,
\\ \nonumber
Q_\nu=\biggl(\frac{p_-^{'}}{p_-^{'}k_1}-\frac{p_-}{p_-k_1}\biggr)_\nu \, ,
\end{gather}
which obeys the gauge conditions
\begin{gather}
E^{born}_{\mu\nu\rho}k_{1\nu}=0, \quad
E^{born}_{\mu\nu\rho}q_{\mu}=0, \quad
E^{born}_{\mu\nu\rho}q_{1\rho}=0.
\end{gather}

\section{1-loop virtual corrections}

\begin{figure} 
\includegraphics[scale=1]{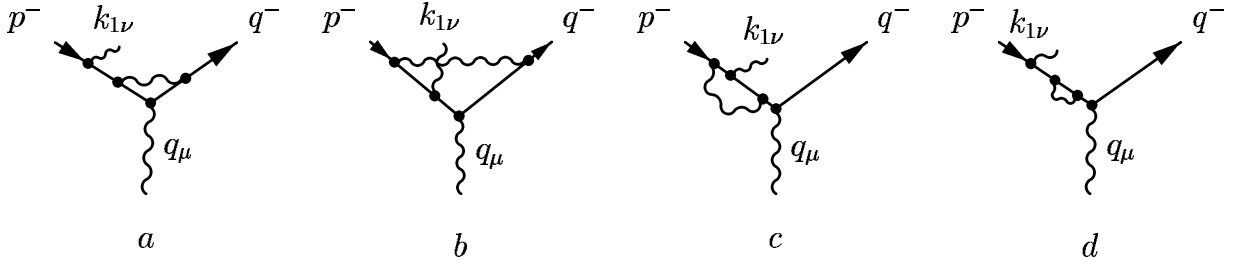}
\caption{Some 1-loop FD for virtual Compton scattering.}
\label{fig2}
\end{figure}

In LLA $\frac{\alpha}{\pi}\ln\frac{q^2}{m_e^2}\sim 1$, $\alpha\ll1$
only FD with single photon exchange between hadron and electron blocks 
(see fig.\ref{fig2}) contribute to cross-section. 
In our consideration we omit the FD with two virtual exchanged photons.
In our paper \cite{eemumugamma}, \cite{eeeegamma}   we demonstrate the cancellation
of all such a contributions including soft and hard photon emission
amplitude exchanged between electron and muon blocks.

In real calculations we can consider only  FD  shown at fig.\ref{fig2}.
The total contribution to DVCS tensor can be restored from
the interference of these amplitudes (fig.\ref{fig2}) with Born one (fig.\ref{fig1}c or d):
\begin{gather}
E_{\mu\nu\rho}^{virt}=(1-P(p_-\to-p_-^{'},p_-^{'}\to-p_-))M_{\mu\nu}^\gamma(M_\rho)^\star
\end{gather}

Matrix element describing electron self-energy insertion (see fig.\ref{fig2}c,d)
and vertex function of real photon emission by initial electron have a form:
\begin{gather}
\frac{\alpha}{2\pi}\overline{u}(p_-^{'})\gamma_\mu\biggl[A_1\biggl(\hat{e}-\hat{k_1}\frac{ep_-}{k_1p_-}\biggr)
+A_2\hat{k_1}\hat{e}\biggr]u(p_-).
\end{gather}
Contribution of structure $A_1$ disappears in the limit $m_e\to0$ \cite{tables},
whereas $A_2$ survives, providing the contribution to DVCS tensor:
\begin{gather}
E_{\mu\nu\rho}^{virt_1}=8\frac{\alpha}{2\pi}\frac{1}{\chi_-}(\ln\frac{\chi_-}{m_e^2}-\frac{1}{2})\frac{1}{4}
Sp\,\, \hat{p}_-^{'}\hat{\gamma}_\mu\hat{k_1}\hat{\gamma}_\nu\hat{p}_-\hat{\gamma}_\rho\, ,
\end{gather}
Contributions of virtual photon emission vertex type FD (fig.\ref{fig2}a) as well as box-type (fig.\ref{fig2}b)
have a form:
\begin{gather}
E_{\mu\nu\rho}^{virt_2}=2\frac{\alpha}{\pi}\int\frac{\dd^4k}{i\pi^2}\biggl\{\frac{S_1}{\chi_-}+\frac{S_2}{(p_--k)^2-m_e^2}\biggr\}
\frac{1}{(k^2-\lambda^2)((p_--k)^2-m^2)((p_--k1-k)^2-m^2)}
\end{gather}
where
\begin{gather}
S_1=\frac{1}{4}
Sp\,\, \hat{p}_-^{'}\hat{\gamma}_\lambda(\hat{p}_-^{'}+\hat{k})\hat{\gamma}_\mu
(\hat{p}_-^{'}-\hat{k}_1-\hat{k})\hat{\gamma}_\lambda(\hat{p}_--\hat{k})\hat{\gamma}_\nu
\hat{p}_-\hat{\gamma}_\rho\, , 
\\ \nonumber
S_2=\frac{1}{4}
Sp\,\, \hat{p}_-^{'}\hat{\gamma}_\lambda(\hat{p}_-^{'}+\hat{k})\hat{\gamma}_\mu
(\hat{p}_-^{'}-\hat{k}_1-\hat{k})\hat{\gamma}_\nu(\hat{p}_--\hat{k})\hat{\gamma}_\lambda
\hat{p}_-\hat{\gamma}_\rho\, .
\end{gather}
Their calculation require scalar, vector and tensor  (up to three rank)
integrals with three and four denominators, listed in \cite{tables}.

The sum of all vertex contributions except only FD with Dirac form-factor (with logarithmical approximation) are:
\begin{gather}
E_{\mu\nu\rho}^{virt}=E_{\mu\nu\rho}^{virt_1}+E_{\mu\nu\rho}^{virt_2}=E_{\mu\nu\rho}^{0}(-\frac{1}{4}\ln^2\frac{-t}{m_e^2}
+\frac{1}{2}\ln\frac{m_e^2}{\lambda^2}(\ln\frac{-t}{m_e^2}-1)
-\frac{3}{4}\ln\frac{-t}{m_e^2}).
\end{gather}
In this expression was used the assumption that all term proportional to
$k_{1\nu}$ is equal to zero. It is validate due to real photon $k_1$,
for which $k_1^2=0$. 

\section{soft photon and Dirac form-factor contributions}
Finally we must consider the vertex-type corrections to electron scattering vertex
without real photon emission and the contribution of additional
soft photon emission with energy nit exceeded $\Delta\varepsilon$

Both these contributions are proportional to Born DVCS terms:
\begin{gather}
E_{\mu\nu\rho}^{soft+D}=E_{\mu\nu\rho}^{0}\biggl(\frac{\alpha}{\pi}\Gamma_1(q_1^2)+\delta_{soft}\biggr),
\\ \nonumber
\delta_{soft}=\frac{4\pi\alpha}{(2\pi)^3}\int\frac{\dd^3k_2}{\omega_2}
\biggl(\frac{p_-}{2p_-k_2}-\frac{p_-^{'}}{2p_-^{'}k_2}\biggr)^2\biggl|_{\omega_2\ll\Delta\varepsilon}
\end{gather}
where
\begin{gather}
\frac{\alpha}{\pi}\Gamma_1(q_1^2)=\frac{\alpha}{\pi}
\biggl[(\ln\frac{m_e}{\lambda})(1-\ln\frac{-q_1^2}{m_e^2})-\frac{1}{4}\ln^2\frac{-q_1^2}{m_e^2}
-\frac{1}{4}\ln\frac{-q_1^2}{m_e^2}+\frac{\pi^2}{12}\biggr],
\\ \nonumber
\delta_{soft}=-\frac{\alpha}{\pi}
\biggl[(\ln\frac{-q_1^2}{m_e^2}-1)\ln\frac{(\Delta\varepsilon)^2m_e^2}{\lambda^2\varepsilon_-\varepsilon_-^{'}}
+\frac{1}{2}\ln^2\frac{-q_1^2}{m_e^2}-\frac{1}{2}\ln^2\frac{\varepsilon_-^{'}}{\varepsilon_-}
-\frac{\pi^2}{3}+\mathrm{Li}_2(\cos\frac{\theta}{2})\biggr],
\end{gather}
Here $\theta$ is electron scattering angle, $\varepsilon_-$, $\varepsilon_-^{'}$
are energies  of initial and scattered electrons (in cms of electron and proton).

Combining all contributions containing large logarithms,
we arrive to the lowest order expansion of r.h.c. which does not contain auxiliary
parameter $\lambda$.  General
statement about renormalization group approach and 
the Drell-Yan picture of cross-section provides the validity of our
result in LLA at all orders of perturbation theory.
\begin{gather}
E_{\mu\nu\rho}^{summed}=E_{\mu\nu\rho}^{virt}+E_{\mu\nu\rho}^{soft+D}=E_{\mu\nu\rho}^{0}
(\ln\frac{(\Delta\varepsilon)^2}{\varepsilon_-\varepsilon_-^{'}}(\ln\frac{-t}{m_e^2}-1)
+6\ln\frac{-t}{m_e^2}).
\end{gather}

\section{additional hard photon contribution}

Contributions arising from additional hard photon emission with energy $\omega_2>\Delta\varepsilon$
can be put in two parts. One, corresponding to collinear kinematics,
contain a large logarithm of type $\ln\frac{-t}{m_e^2}$
can be found using the quasireal electron method \cite{BFK}.
It has a form:
\begin{gather}
\frac{\alpha}{\pi}\int\limits_{0}^{1-\Delta_1}\dd x P(x)\ln\frac{-t}{m_e^2}E_{\mu\nu\rho}^{0}(p_-x,p_-^{'},k_1),
\end{gather}
for the case of photon emission close to initial electron, and
\begin{gather}
\frac{\alpha}{\pi}\int\limits_{y}^{1-\Delta_2}\frac{\dd z}{z} P(\frac{y}{z})\ln\frac{-t}{m_e^2}
E_{\mu\nu\rho}^{0}(p_-,\frac{z}{y}p_-^{'},k_1),
\end{gather}
 for the case of photon emission close to scattered electron with
\begin{gather}
\Delta_1=\frac{\Delta\varepsilon_1}{\varepsilon_-} , \quad 
\Delta_2=\frac{\Delta\varepsilon_2}{\varepsilon_-^{'}}, \quad
P(z)=\frac{1+z^2}{1-z}.
\end{gather}
Contribution from noncollinear kinematics does not contain
large logarithms.

\section{Drell-Yan picture of the process}

By summing up all contributions , we can put the
cross section of the radiative production in the form
of Drell-Yan process:
\ba \label{18}
&& E_{\mu\nu\rho}(p_-,p_-^{'},k_1)
= \int\limits_{0}^{1}\dd x D(x)\int\limits_{y}^{1} \frac{\dd z}
{z}D(\frac{y}{z})
\frac{E_{\mu\nu\rho}^{0}(xp_-,\frac{z}{y}p_-^{'},k_1)}
{q^2(x,z)q_1^2(x,z)}
\\ \nonumber &&
D(x)=\delta(1-x)+\frac{\alpha}{2\pi}P^{(1)}(x)(\ln\frac{-t}{m_e^2}-1)+....,
\; 
\\ \nonumber
&&P^{(1)}(x)=\lim_{\Delta\to0}\biggl[(\ln\frac{\Delta\varepsilon}{\varepsilon}+3)\delta(1-z)
+\Theta(1-z-\Delta)\frac{1+z^2}{1-z}\biggr]
=\biggl(\frac{1+x^2}{1-x}\biggr)_+ \,\, .
\ea
Here we define
\begin{gather}
q(x,z)=xp_--\frac{z}{y}p_-^{'}, \quad q_1(x,z)=q(x,z)-k_1.
\end{gather}
\section{Conclusion}

We calculate the VCS for high-energy limit. Also we include 
additional hard photon for the case of collinear emission.
Sum of all contributions (including soft photon emission)
does not depend on fictitious photon mass $\lambda$
and is consistent with renorm group prediction.

As a realistic hadronic tensor we can consider the case of muon-electron radiative scattering, 
which can be naive model for quark-electron scattering:
\begin{gather} 
e_-(p_-)+\mu_-(q_-)\to e_-(p_-^{'})+\mu_-(q_-^{'})+\gamma(k_1),
\end{gather}
By combining our DVCS tensor with "quark" one
\begin{gather}
\label{19}
H_{\mu\nu\rho}=
Sp\,\, \hat{q}_-^{'}(Q_\nu\hat{\gamma}_\mu+\frac{\hat{\gamma}_\mu\hat{k_1}\hat{\gamma}_\nu}{\tilde{\chi}_-}
+\frac{\hat{\gamma}_\nu\hat{k_1}\hat{\gamma}_\mu}{\tilde{\chi}_-^{'}})\hat{q}_-\hat{\gamma}_\rho\, ,
\\ \nonumber
Q_\nu=\biggl(\frac{q_-^{'}}{q_-^{'}k_1}-\frac{q_-}{q_-k_1}\biggr)_\nu \, ,
\\ \nonumber
H_{\mu\nu\rho}k_{1\nu}=0, \quad H_{\mu\nu\rho}q_{1\mu}=0, \quad H_{\mu\nu\rho}q_{1\rho}=0,
\end{gather}
the relevant  DVCS contribution to summed square of matrix element are 
(see fig.\ref{fig1}a,b,c,d, we consider only interference between FD with 
emission from muon and electron blocks):
\begin{gather}
\Delta|M|^{2 (e\mu\gamma)}_{DVCS}=\frac{16(4\pi\alpha)^3}{tt_1}(s^2+s_1^2+u^2+u_1^2)
\biggl[\frac{s}{\chi_-\chi_-^{'}}+\frac{s_1}{\tilde{\chi}_-\tilde{\chi}_-^{'}}
+\frac{u}{\tilde{\chi}_-^{'}\chi_-}+\frac{u_1}{\chi_-^{'}\tilde{\chi}_-}\biggr], \\ \nonumber
(\dd\sigma)^{e\mu\gamma}_{DVCS}=\frac{1}{8s}\frac{\Delta|M|^{2 (e\mu\gamma)}_{DVCS}}
{(1-\Pi(t))(1-\Pi(t_1))}\dd\Gamma,
\end{gather}
and here 
\begin{gather}
\dd\Gamma=\frac{\dd^3p_-^{'}}{2\varepsilon_{e_-^{'}}}\frac{\dd^3q_-^{'}}{2\varepsilon_{q_-^{'}}}
\frac{\dd^3k}{2\omega}\frac{\delta^4(p_-+q_--p_-^{'}-q_-^{'}-k_1)}{(2\pi)^5}.
\end{gather}
Formula (\ref{18}) combining with (\ref{19}) for ${e\mu\to e\mu\gamma}$ process
is consistent with corresponding part of Drell-Yan picture formula
in paper \cite{eemumugamma}, the whole part of which included also all contributions from "box" diagram and
photon emission from muon line.

\end{document}